# МАТЕМАТИЧЕСКИЙ И ПРОГРАММНЫЙ АППАРАТ АВТОМАТИЗИРОВАННОГО ПРОЕКТИРОВАНИЯ АКСОНОМЕТРИЧЕСКИХ СХЕМ ТРУБОПРОВОДНЫХ СИСТЕМ

В.В. Мигунов

Рассматривается применение модульной технологии разработки проблемно-ориентированных расширений систем автоматизированного проектирования [1] к задаче автоматизации подготовки аксонометрических схем трубопроводных систем (АСТС). В [2] выявлено единство состава схем для специальных технологических трубопроводов, систем водопровода и канализации, отопления, теплоснабжения, вентиляции, кондиционирования воздуха и приведено структурированное параметрическое представление схем, включая свойства объектов и их связи, общие установки и установки по умолчанию. В [3] описаны особенности реализации типовых операций, состав и реализация специальных операций проектирования.

Реализация плоских схем трехмерных каркасных объектов (принято присваивать таким моделям размерность 2,5) ввиду относительно большого объема вычислений требует специализированного модельного, методического и математического подходов, в общем случае отсутствующих в двумерных графических редакторах САПР. Эти подходы, реализованные в САПР TechnoCAD GlassX, излагаются ниже.

*Системы координат*

Во всех системах координат единицей измерения являются миллиметры. В самом чертеже имеются системы координат Натура (Re) и Бумага (Pr), отличающиеся масштабом всего чертежа, указываемом в основной надписи. У экранной системы координат Sc ось X идет вправо, Y - вниз, единица измерения - пиксел. Специально для АСТС вводятся дополнительные декартовы системы координат.

ReIn - правая трехмерная. Соответствует представлению концов труб в реальном трехмерном пространстве. Начало координат задается точкой привязки АСТС в чертеже. Единица измерения - мм Натуры.

ReVi - видимая трехмерная. Соответствует представлению концов труб и всех других элементов АСТС в трехмерном пространстве после всех смещений. Во всем совпадает с ReIn, но расстояния в ней не соответствуют реальным длинам, если действуют смещения. Смещения в АСТС с условным изображением длины труб разрешены соответствующими ГОСТами. Общие смещения действуют на полупространства, местные - на ветки труб, идущие от одной или нескольких параллельных труб.

Li - библиотечная, двумерная. Соответствует представлению блоков (условных графических обозначений трубопроводной арматуры и элементов трубопроводов) в графической библиотеке АСТС. Начало координат является точкой привязки блока, ее координата вдоль трубы задается при привязке блока к трубе. Ось X является осью привязки блока к трубопроводу, совпадает с осью трубы после привязки. Вдоль этой оси задается отрезок, вырезаемый блоком на трубе при установке на нее. Единица измерения - мм Бумаги.

*Типы данных*

Используются четырехбайтные вещественные числа (Float), что продиктовано требованием компактности хранения АСТС. Вводятся типы данных: V2, V3 - векторы из 2 и 3 компонентов Float; M22, M32, M23, M33 - матрицы с соответствующим числом строк и столбцов из элементов Float.

В терминах таких типов задаются направленные отрезки (V2, V2) в Li, Re, Pr, Sc и (V3, V3) в ReIn, ReVi; ориентированная плоскость в ReIn (V3, Float) - орт нормали в положительном направлении и координата плоскости на этом направлении; оператор проецирования (V2, M23) из ReIn и ReVi в Re, Pr, Li, Sc - линейное преобразование и сдвиг; аналогичные операторы размещения блока (V3, M32) из Li в ReIn и оператор перехода (V2, M22) из Li в Re, Pr, Sc.

*Общие принципы реализации операций над данными*

Состав и реализация операций над такими данными подчинены требованиями компактности и наглядности использующего их программного кода, а также быстродействия. Во всех случаях передача аргументов длиной более 4 байтов осуществляется по ссылке, а не по значению. Для использования одних и тех же процедур для векторов и строк/столбцов матриц в описании интерфейса аргументы указываются как нетипированные переменные.

*Состав основных операций*

Чтение и запись значений Float, расположенных в элементе массива с заданным номером.

Проверки с заданной погрешностью равенства нулю и единице чисел Float, покомпонентного равенства векторов V2, V3.

Обнуление, смена знака, сложение и вычитание векторов V2 и V3.

Умножение векторов V2 и V3 на скаляр, умножение на скаляр заданного числа идущих подряд элементов Float.

Скалярные произведения векторов V2 и V3. Скалярное произведение X: V2 на нормаль к Y: V2, получаемую путем поворота Y на 90° против часовой стрелки. Модуль этого произведения равен расстоянию от конца X до линии, продолжающей Y, если Y - орт.

Вычисление длин и ортов векторов V2 и V3, расстояний между точками V2 и V3 и ортов направлений от одной к другой.

Векторное умножение векторов V3.

Определение скаляра k в формуле Y = k * X для коллинеарных X, Y: V3 с заданной погрешностью.

Определение с заданной погрешностью индекса оси (1 - +X, 2 - +Y, 3 -

+Z, -1 - -X, -2 - -Y, -3 - -Z), в направлении которой идет вектор V3. Обратная операция - вычисление орта оси по ее индексу и орта трубы по ее номеру в списке. Эта довольно необычная операция нужна для выяснения ориентации труб при нанесении цепных размеров в ReVi, и соответственно, при установке блоков на трубы. Дело в том, что выносные линии должны находиться в плоскости блока и одновременно идти вдоль одной из осей координат. Допустимый набор ориентации плоскостей блоков и размеров зависит от взаимного расположения трубы и осей координат ReIn.

Умножение векторов на матрицы: M22*V2, M23*V3, M32*V2, M33*V3, M22*V2.

Умножение матриц M23*M32 => M22.

Вычисление определителя матрицы M22.

Пересчеты координат точек V3 при переходах между системами координат ReIn и ReVi с учетом общих, местных и всех смещений.

Пересчет координат точек V3 из ReVi в систему координат Бумага чертежа.

Проверка того, что матрица M23 проецирования из ReVi в Pr превращает плоскость, образуемую двумя непараллельными ненулевыми векторами M3Rec, в прямую линию.

Приведенные сведения могут быть полезны при разработке программных средств, оперирующих в пространстве размерности 2,5.

Литература
1. Мигунов В.В. Модульная технология разработки проблемно-ориентированных расширений САПР реконструкции предприятия / Материалы Второй международной электронной научно-технической конференции "Технологическая системотехника" (ТСТ'2003), г.Тула, 01.09.2003-30.10.2003 [Электронный ресурс] / Тульский государственный университет. – Режим доступа: http://www.tsu.tula.ru/aim/, свободный. – Загл. с экрана. – Яз. рус., англ.
2. Мигунов В.В., Кафиятуллов Р.Р., Сафин И.Т. Модульная технология разработки расширений САПР. Аксонометрические схемы трубопроводов. Параметрическое представление / Там же.
3. Сафин И.Т., Мигунов В.В., Кафиятуллов Р.Р. Модульная технология разработки расширений САПР. Аксонометрические схемы трубопроводов. Типовые и специальные операции / Там же.

Мигунов Владимир Владимирович; к.т.н.; ст. науч. сотр.; член-корр. Международной академии информатизации; нач. отдела инф-х технологий Центра экономических и социальных исследований Республики Татарстан; 420088 Казань, Губкина, д.50; (8432) 73-37-14; vmigunov@csp.kazan.ru.